\documentclass[aps,pra,10pt,twocolumn,showpacs,superscriptaddress,footnoteinbib]{revtex4}
\usepackage{amsmath}
\usepackage{latexsym}
\usepackage{amssymb}
\usepackage{color}
\usepackage{graphicx}
\usepackage[outdir=./]{epstopdf}
\usepackage{graphics}
\usepackage{soul}
\usepackage[breaklinks]{hyperref}

\begin{document}

\title{Complementarity Relations Between Quantum Steering Criteria}
\author{Debasis Mondal}
\thanks{cqtdem@nus.edu.sg} 
\affiliation{Centre for Quantum Technologies, National University of Singapore, 3 Science Drive 2, Singapore 117543}
\author{Dagomir Kaszlikowski}
\thanks{phykd@nus.edu.sg} 
\affiliation{Centre for Quantum Technologies, National University of Singapore, 3 Science Drive 2, Singapore 117543}
\affiliation{Department of Physics, National University of Singapore, 2 Science Drive 3, 117542 Singapore, Singapore}
\date{\today}

\begin{abstract}
 
Recently, a connection between quantum coherence and quantum steering was established and criteria for quantum steering or in other words, nonlocal advantage of quantum coherence (NAQC) were derived for two-qubit states. Here, we derive a set of complementarity relations between the steering or NAQC inequalities  achieved by various criteria. We also extend the idea in the multi-partite scenario, specifically, in the three-qubit scenario, which can easily be generalized to the multi-partite scenario.
\end{abstract}


\maketitle
\section{Introduction.}Recently, complementarity and trade-off relations in different forms for different quantities and to establish different notions of quantum information theory have turned into a new trend. To name a few, we have coherence complementarity relations \cite{hall, pan, deba1}, information complementarity for decoding quantum incompatibility \cite{zhu}, contextuality-nonlocality trade-off relation \cite{Zhanx}, coherence-mixedness complementarity relation \cite{uttam1} and reverse uncertainty relations \cite{deba2}, which is nothing but a new complementarity relation between variances of two quantum mechanical observables. These relations are important in the sense that they establish a bound on a quantity through another complementary quantity and also helps to understand the geometry of the quantum state space and information as well as correlation.

So far, several steering inequalities have been derived on the basis of the existence of single system description of a part of the bi-partite systems~\cite{Saunders, FUR_St, Walborn}. It has also been quantified for two-qubit systems \cite{EPR_quant}.
In the last few years, several experiments have been performed to demonstrate the steering effect with the increasing number of measurement settings~\cite{Saunders} and with loophole free arrangements~\cite{Exp_St_1}. For continuous variable systems, the steerability has also been quantified \cite{gerardo}.

Quantum superposition, on the other hand, is one of the main reasons for quantum mechanical advantages in quantum computation, quantum cryptography, teleportation and dense coding and behind all the exclusively quantum mechanical resources. There was always a need to measure and quantify quantum superposition and that was finally fulfilled with the publication of quantifying coherence \cite{l1_norm} and \cite{girolami,aberg06}. After its introduction into the literature, it has turned into a trendy topic, particularly to understand its connection with quantum entanglement, Bell nonlocality, other resource theories  \cite{deba, spekken} and even as a resource in quantum thermodynamics \citep{guzic,gour,rudolph1,rudolph2,gardas}.

Understanding the relations between various resources and their underlying geometries have utmost importance in quantum information theory. Quantum steering or EPR nonlocality is one such resource \cite{aolita} and it is natural to be curious about whether it may be employed to control the coherence of a system nonlocally. It turned out that it is indeed possible. An intriguing connection between quantum coherence and quantum steering was established \cite{deba1}. In this paper, we dig deeper into the topic and try to understand the underlying geometry of quantum steering and how it affects the coherence of a part of the system nonlocally.

We derive a set of complementarity relations between various criteria for achieving nonlocal advantage of quantum coherence (NAQC). We are already aware of monogamy relation for entanglement \cite{koashi,dong}, Bell \cite{debsh,paterek,qip} and EPR nonlocality \cite{reid,rudolph}, which is nothing but an inequality between entanglement or nonlocality between three parties. On the other hand, complementarity relation in this paper is an inequality between various settings on two fixed parties. We show that if one steering inequality is violated, the inequality with complementary settings cannot be violated by the state.

\begin{figure}
\includegraphics[scale=.4]{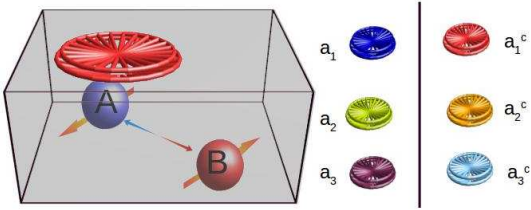}
\caption{Graphical depiction of a table of complementary keys (``steerings") $(a, a^c)$ to access some quantum resource (NAQC) for quantum information theoretic protocols. If one key works, the other complementary one does not. Resource cannot be accessed even if one tries to use both of the complementary keys simultaneously.}
\label{fig1comp}
\end{figure}

To visualize, let us consider a classical scenario. We consider a set of complementary pairs of keys (``steerings"), as depicted in the Fig. (\ref{fig1comp}). Our aim is to open and access some resources from the box using these keys. We call these pairs of keys complementary for the following reasons:  If one uses both the keys of a pair one after another to unlock the box, it cannot be opened. If they are used separately one by one only one key of the pairs can open and access the resource. That means the other complementary key must compensate the result of another so that they together cannot achieve this advantage. Now, this classical model cannot be taken literally for our quantum mechanical phenomenon. But at least, it is now clear from this example that this work classifies a set of steering inequalities. This classification should be helpful in understanding the geometry of how to extract the extra quantum coherence due to nonlocality and use that as a resource in various information theoretic protocols.

In \cite{deba1} (see supplemental material), a connection between the speed of quantum evolutions and the effect of quantum nonlocality on it was established. This work will surely help us to control and manipulate the speed of quantum system or engines in quantum thermodynamics using quantum nonlocality.

At first, we generalize and derive a set of NAQC or steering inequalities following \cite{deba1} based on different measurement settings. Violation of anyone of them implies NAQC or as has been argued in \cite{deba1}, a sign of steerability. We classify these criteria based on their measurement settings and in the next section, we establish a set of complementarity relations between them in the sense that if one particular criterion is violated and the state turns out to be a steerable state, the other complementary criterion cannot be violated. We also generalize the steering or NAQC inequalities in the three-qubit scenario based on these two-qubit complementarity relations and at last, in turn, use
these inequalities to show the three-qubit steering or NAQC complementarity inequalities. 
\section{Steering Inequalities}
Here, we follow the same protocol as steering as was first laid down in \cite{Jones07} and later taken up in \cite{deba1} in the context to show the advantage in quantum coherence due to nonlocality. We consider a hypothetical game, where Alice prepares two quantum systems, say, $A$ and $B$ in an entangled state $\rho_{AB}$ and sends the system $B$ to Bob. Bob does not trust Alice but agrees with the fact that the system $B$ is quantum. Therefore, Alice's task is to convince Bob that the prepared state is indeed entangled and they share nonlocal correlation. On the other hand,  Bob thinks that Alice may cheat by preparing the system $B$ in a single quantum system, on the basis of possible strategies~\cite{Saunders, FUR_St}. Bob agrees with Alice that the prepared state is entangled and they share nonlocal correlation if and only if the state of Bob cannot be written by local hidden state model (LHS)~\cite{Jones07}
\begin{eqnarray}
\rho^A_a = \sum_{\lambda} \mathcal{P}(\lambda)\,\mathcal{P}(a|A,\lambda)\,\rho_{B}^Q(\lambda),
\label{LHS}
\end{eqnarray}
where $\{\mathcal{P}(\lambda),\rho_{B}^{Q}\}$ is an ensemble of pre-existing local hidden states of Bob and  $\mathcal{P}(a|A,\lambda)$ is Alice's stochastic map to convince or fool Bob by preparing a state $\rho^{A}_{a}$.
Here, we consider $\lambda$ to be a hidden variable with the constraint $\sum_{\lambda}\mathcal{P}(\lambda)=1$ and $\rho_{B}^{Q}(\lambda)$ is a quantum state received by Bob. 
The joint probability distribution on such states, $P(a_{\mathcal{A}_i},b_{\mathcal{B}_i})$ of obtaining outcome $a$ for the measurement of observables chosen from the set $\{\mathcal{A}_i\}$ by Alice and outcome $b$ for the measurement of observables chosen from the set $\{\mathcal{B}_i\}$ by Bob can be written as
\begin{eqnarray}
P(a_{\mathcal{A}_i},b_{\mathcal{B}_i}) = \sum_{\lambda} P(\lambda)\, P(a_{\mathcal{A}_i}|\lambda)\, P_Q(b_i|\lambda),
\label{LHS_JP}
\end{eqnarray}
where $P_Q(b_i|\lambda)$ is the quantum probability of the measurement outcome $b_{i}$ due to the measurement of $\mathcal{B}_{i}$.

To detect the steerability, a set of complementarity relations between coherences measured in mutually unbiased bases were set up \cite{deba1} for various measures of quantum coherence. For an arbitrary qubit state $\rho$, these relations can be written as
\begin{equation}\label{cohcomp}
\sum_{i}C_{i}^{q}\leq \epsilon^{q},
\end{equation}
where  throughout the paper we consider $q\in\{l_1, E, S\}$ depending on various measures of quantum coherence, $l_1-$norm $(l_1)$, relative entropy of coherence $(E)$ and skew-information (S) such that $\epsilon^q\in \{\sqrt{6}, 2.23, 2\}$ respectively and $i\in\{1, 2, 3\}$ depending on the three mutually unbiased bases of Pauli matrices $\{\sigma_1, \sigma_2, \sigma_3\}$ respectively.

Let us now describe our steering protocol, which we use to observe the effects of steering on the coherence of a part of a bi-partite system. We consider a general two-qubit state of the form of
\begin{eqnarray}
\eta_{AB}= && \frac{1}{4} (I^A\otimes I^B+\vec{ r} \cdot {\sigma}^{A}\otimes I^B+  I^{A}\otimes \vec{s}\cdot \vec{\sigma}^{B} \nonumber \\
&&+\sum_{i,j=x,y,z} t_{ij} \sigma_{i}^{A}\otimes \sigma_{j}^{B}),
\label{G_2qbit}
\end{eqnarray}
where $\vec{r}\equiv(r_{x},r_{y},r_{z})$, $\vec{s}\equiv(s_{x},s_{y},s_{z})$, 
with $|r|\leq 1$, $|s|\leq 1$ and $(t_{ij})$ is the correlation matrix.

  Alice may perform measurements in arbitrarily chosen bases. For simplicity, we derive the coherence steerability criteria for three measurement settings by Alice in the eigenbases of Pauli matrices $\{\sigma_{1}, \sigma_{2}, \sigma_{3}\}$. 
 When Alice declares that she performed measurement on one of the three eigenbases of Pauli matrices and obtains outcome $a\in\{0,1\}$ with probability $\{p(\eta_{B|\Pi_{i}^a})=Tr[(\Pi^{a}_{i}\otimes I_{B})\eta_{AB}]|i\in 1, 2, 3\}$, Bob measures coherence with respect to the eigenbasis of any one or two of the three Pauli matrices on his conditional state $\eta_{B|\Pi_i^a}$ (normalised), where $\Pi_{i}^a$ denotes projection operator in the eigenbasis of the pauli matrix $\sigma_{i}$ with outcome $a\in \{0, 1\}$. This procedure is repeated for all of the Alice's measurements in three mutually unbiased bases. Thus, we classify the coherence steering inequalities in two classes based on the number of coherence measurements by Bob for each projective measurements performed by Alice and refer them as one measurement setting and two measurement setting respectively. We show that no state with local hidden state (LHS) model but only other states can violate these criteria. We also show that in case of three measurements of coherence by Bob in the three bases for each projective measurements by Alice, no state can violate the inequality. This leads to a complementarity relations between steering criteria.

To start with, let us first describe the one measurement setting and then two measurement setting will follow from the result. We consider Bob to measure the coherence only in one Pauli basis for each projective measurements performed by Alice. For simplicity, let us consider that Alice and Bob both perform the measurement in the same basis. We know that Alice's measurement in a particular basis affects the coherence of Bob's conditional state. We define a quantity 

\begin{equation}
S^B_{j}(\rho_{AB})=\sum_{i, a}p(\rho_{B|\Pi_{i}^{a}})C_{i+j}^{q}(\rho_{B|\Pi_i^a}),
\end{equation}
which is a probabilistic combination of coherences of Bob's conditional states. Here, upper index on $S$, $1$ denotes single measurement setting and $j\in\{0, 1, 2\}$.
If the conditional state $\rho_{B|\Pi_{k}^a}$ of bob is considered to be prepared by a local hidden model prepared by Alice, i.e., $\rho_{B|\Pi_{k}^a}=\frac{\rho^{\Pi_{i}}_{a}}{p(\rho^{\Pi_{i}}_{a})}$ with probability $p(\rho^{\Pi_i}_{a})$ as given by Eq. (\ref{LHS}), then from the definition of LHS as given in Eq. (\ref{LHS}) we get
\begin{eqnarray}\label{s1}
S^B_{0}(\rho_{AB})&=&\sum_{i, a}p(\rho^{A}_{a})C_{i}^{q}\bigg(\frac{\sum_{\lambda}\mathcal{P}(\lambda)\mathcal{P}(a|\Pi_{i},\lambda)\rho_{B}^{Q}(\lambda)}{p(\rho^{\Pi_{i}}_{a})}\bigg)\nonumber\\&\leq & \sum_{i, a, \lambda}\mathcal{P}(\lambda)\mathcal{P}(a|\Pi_i, \lambda)C_{i}^{q}(\rho^{Q}_{B}(\lambda))\nonumber\\&=&\sum_{i, \lambda}\mathcal{P}(\lambda)C_{i}^{q}(\rho^{Q}_{B}(\lambda))\nonumber\\&\leq &\sum_{\lambda}\mathcal{P}(\lambda)\epsilon^{q}=\epsilon^{q},
\end{eqnarray}
where in the last inequality we use the coherence complementarity relation as stated in Eq. (\ref{cohcomp}) and $\rho^{Q}_B$ is a quantum state received by Bob following local hidden state model. We arrived at the forth line equality by taking summation over $a$ on the quantity in the third line. This is the one setting steering criterion. Similarly, one could derive the two setting steering inequality as proposed in \cite{deba1} as

\begin{eqnarray}
\sum_{\substack{i,j,a}}p(\rho_{B|\Pi_{j\neq i}^a})C_i^{q}(\rho_{B|\Pi_{j\neq i}^a})\leq 2\epsilon^q,
\end{eqnarray}
which can be represented as
\begin{equation}\label{s2}
S^B_{12}(\rho_{AB})=S^B_{1}(\rho_{AB})+S^{B}_{2}(\rho_{AB})\leq 2\epsilon^q.
\end{equation}
In general, one could generalize the idea and propose a set of coherence steering inequalities as
\begin{equation}
S^{B}_{jk}(\rho_{AB})=S^{B}_{j}(\rho_{AB})+S^B_k(\rho_{AB})\leq 2\epsilon^q,
\end{equation}
where we consider $j\neq k$ and $j, k \in\{0, 1, 2\}$.
These two inequalities are valid for any state with LHS model and a violation by a quantum state like $\eta_{AB}$ implies that the state is not only steerable but also can achieve nonlocal advantage of quantum coherence. 

Now, this is the ideal time to discuss why three measurement settings cannot be considered for detecting steerability of a state. In fact one could define a third kind of inequality using three measurement settings in the following way
\begin{equation}
S^{B}_{jkl}(\rho_{AB})=S_{j}^B(\rho_{AB})+S_{k}^B(\rho_{AB})+S_{l}^B(\rho_{AB})\leq 3\epsilon^q,
\end{equation}
where we always consider $j\neq k\neq l$ and $j, k, l\in\{0, 1, 2\}$. Thus, it turns out to be a single inequality, i.e., 
\begin{equation}
S^B_{012}(\rho_{AB})\leq 3\epsilon^q.
\end{equation}

\section{Complementarity Relations}
In this section, we show that the inequality is trivially satisfied by all the 2-qubit quantum states or in other words, there is no state which can violate this inequality. This new inequality will turn out to be a crucial one in depicting the complementarity relations between various steering criteria as given in Eq. ({\ref{s1}), (\ref{s2}).
\begin{figure}
\includegraphics[scale=.85]{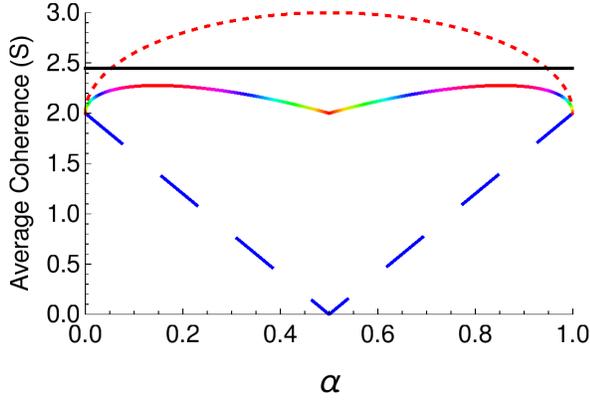}
\caption{We plot the right hand side ($S$) of $S^{B}_{0}(\rho_{AB})\leq \epsilon^{l_1}$ (blue, dashed), $\frac{1}{2}S^{B}_{12}(\rho_{AB})\leq \epsilon^{l_1}$ (red, dotted) and $\frac{1}{3}S^{B}_{012}(\rho_{AB})\leq \epsilon^{l_1}$ (multi-colored continuous plot) for the state $|\psi\rangle=\sqrt{\alpha}|00\rangle+\sqrt{1-\alpha}|11\rangle$. Here, black straight line parallel to the $\alpha$-axis shows $S=\epsilon^{l_1}=\sqrt{6}$. It is clear from the plot that the steering inequality $\frac{1}{2}S^{B}_{12}(\rho_{AB})\leq \epsilon^{l_1}$ is violated by the state, whereas, $S^{B}_{0}(\rho_{AB})\leq \epsilon^{l_1}$ compensates so that $\frac{1}{3}S^{B}_{012}(\rho_{AB})\leq \epsilon^{l_1}$ is never violated. }
\label{fig}
\end{figure}
For this, we consider a general two qubit state as given in Eq. (\ref{G_2qbit}). To derive this inequality, we start with
\begin{eqnarray}\label{eall}
S^B_{012}(\rho_{AB})&=&\sum_{i, j, a}p(\rho_{B|\Pi_{i}^{a}})C_{j}(\rho_{B|\Pi_{i}^{a}})\leq\sum_{i, a}p(\rho_{B|\Pi_{i}^{a}})\epsilon^{q}\nonumber\\&=&3\epsilon^{q}.
\end{eqnarray}
From the derivation, it should be cleared that we never used the LHS model of the conditional states of Bob. Conditional states of Bob for various measurements by Alice may or may not have LHS model. The inequalities in Eq. (\ref{s1}) and (\ref{s2}) can be turned into the inequality given in Eq. (\ref{eall}) just by adding them together. The inequality (\ref{eall}) is satisfied by all the 2-qubit quantum states, although (\ref{s1}) and (\ref{s2}) are violated. Thus, the inequality (\ref{eall}) is actually a complementarity relation between the two coherence steering criteria given in Eq. (\ref{s1}) and (\ref{s2}), i.e., 

\begin{equation}
S^B_{012}(\rho_{AB})=S^B_0(\rho_{AB})+S^B_{12}(\rho_{AB})\leq 3\epsilon^q.
\end{equation}
Thus, When one criterion among (\ref{s1}) and (\ref{s2}) is violated, the other one compensates in a way that the inequality in Eq. (\ref{eall}) is never get violated. It turns out that it is not the only way to decompose $S^B_{012}(\rho_{AB})$ quantities defined by $S_1^B$, $S_2^B$ and $S_3^B$. In principle, one could write
\begin{eqnarray}
S^B_{012}(\rho_{AB})&=&S^B_{0}(\rho_{AB})+S^B_{1}(\rho_{AB})+S^B_{2}(\rho_{AB})\nonumber\\&=&S^B_{01}(\rho_{AB})+S^B_2(\rho_{AB})\nonumber\\&=&S^B_{02}(\rho_{AB})+S^B_1(\rho_{AB})\nonumber\\&=&S^B_{12}(\rho_{AB})+S^B_0(\rho_{AB}).
\label{st_comp}
\end{eqnarray}

From these decompositions, it is clear that if one component of the decomposition violates the steering inequality, for example, $S^B_{12}(\rho_{AB})\geq 2\epsilon^q$, the other component(s) of the corresponding decomposition must satisfy the steering inequality, i.e., $S^B_0(\rho_{AB})\leq\epsilon^q$ such that $S^B_{12}(\rho_{AB})+S^B_0(\rho_{AB})\leq 3\epsilon^q$ for any state.

\section{tripartite steering inequality}
There are numerous papers on tripartite steering inequalities \cite{ricardo, wen}, quantification and detection of genuine tripartite entanglement using steering inequalities \cite{natphy, deng, jeba, caval}, authentication protocol using tripartite steerability \cite{deba}. Here we derive a new steering inequality in the tri-partite scenario where one of the parties (say Alice) verify the steerability of the two qubit part (say $A$ and $B$) of the state $\rho_{abc}$ by the other party (say Charlie), where $A$ and $B$ parts of the state belong to Alice and Bob respectively and $C$ belongs to Charlie. To derive the inequality, we start with the LHS model of the two qubit system belonging to Alice
\begin{equation}
\rho^{C}_{c}=\sum_{\lambda}\mathcal{P}(\lambda)\mathcal{P}( c| C, \lambda)\rho_{AB}^{Q}(\lambda).
\label{LHS_model2}
\end{equation}
One way to derive the steering inequality in this situation is to follow the method mentioned in \cite{deba1}, i.e., to derive the coherence complementarity relation for a two-qubit state. Using the coherence complementarity relation, one can easily derive the steering inequality for the steerability of the state belonging to Alice by Charlie($C$). In this paper, we follow another method to achieve the same goal. We consider the steering complementarity relations derived above to establish a new kind of steering inequalities as
\begin{equation}
\sum_{i, c}S^{B}_{i}(\rho_{AB|\Pi_{i}^{c}})\leq 3\epsilon^{q}.
\label{new_steering1}
\end{equation}
Another steering inequality similarly can be 
\begin{equation}
\sum_{i, j, c}S^{B}_{j\neq i}(\rho_{AB|\Pi_{i}^{c}})\leq 6\epsilon^{q}.
\label{new_steering2}
\end{equation}
The main difference between the former and the later method is that in the later method, Alice does not need to own the entire two qubit state to measure coherence on the conditional states. The proof of the steering inequality (\ref{new_steering1}) can be shown following the steering complementarity inequalities given in Eq. (\ref{st_comp}) as
\begin{widetext}
\begin{eqnarray}
\sum_{i, c}S^{B}_{i}(\rho_{AB|\Pi_{i}^{c}})&=&\sum_{c, i}S^{B}_{i}\bigg(\frac{\sum_{\lambda}\mathcal{P}(\lambda)\mathcal{P}(c|i, \lambda)\rho_{AB}(\lambda)}{\sum_{\lambda}\mathcal{P}(\lambda)\mathcal{P}(c|i, \lambda)}\bigg)\nonumber \\& \leq &
\sum_{i, \lambda}S^{B}_{i}(\rho_{AB}(\lambda))
\nonumber \\&  \leq & 3\epsilon^{q}
\end{eqnarray}
\end{widetext}
where we consider $\rho_{AB|\Pi^{c}_{j}}$ is a normalized state with the LHS model as given above in Eq. (\ref{LHS_model2}), i.e., $\rho_{AB|\Pi^{c}_{j}}=\frac{\sum_{\lambda}\mathcal{P}(\lambda)\mathcal{P}(c|i,\lambda)\rho_{AB}^{Q}(\lambda)}{\sum_{\lambda}\mathcal{P}(\lambda)\mathcal{P}(c|i, \lambda)}$. To prove the inequality, we also consider above in the second line that $S(\sum_{i}p_{i}\rho_{AB}^{i})\leq\sum_{i}p_{i}S(\rho^{i}_{AB})$ due to the fact that coherence cannot increase under classical mixing. In Fig. (\ref{fig2}), we show the violation of the inequality in Eq. (\ref{new_steering1})
\begin{figure}
\includegraphics[scale=0.9]{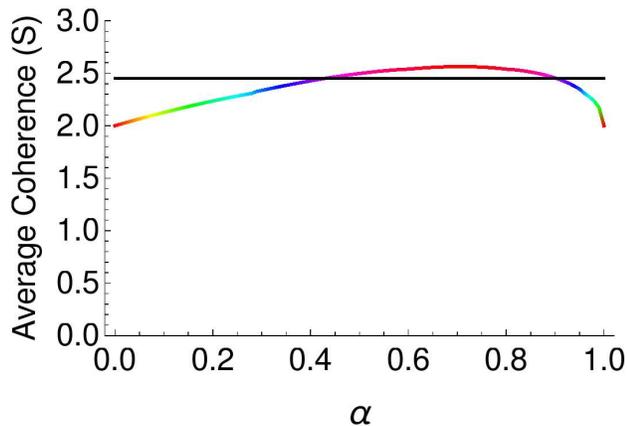}
\caption{We consider a three-qubit $GHZ$ state $|\psi_{GHZ}\rangle=\alpha|000\rangle+\sqrt{1-\alpha^2}|111\rangle$. 
and plot the right hand side ($S$) of the inequality in Eq. (\ref{new_steering1}) with respect to the parameter $\alpha$. The straight line parallel to the $\alpha$-axis is  $3\sqrt{6}$ line. One may argue that the steering inequality is too weak because even GHZ state didn't show the violation for the entire range of parameter $\alpha$. But a counter argument would be that we have not used the free will and only fixed bases, i.e., Pauli bases have been used.}
\label{fig2}
\end{figure}
\section{Tripartite Complementarity relations}We can now use the tripartite steering inequalities given in Eq. (\ref{new_steering1}) and Eq. (\ref{new_steering2}) to derive a set of complementarity steering inequalities for tripartite scenarios as well. One can easily show that the inequality
\begin{equation}
\sum_{i, j, c}S^{B}_{j}(\rho_{AB|\Pi_{i}^{c}})\leq 9\epsilon^{q}
\label{comp2}
\end{equation}
is true for any three qubit systems, i.e., no three qubit state can violate the inequality. Thus, when the first three qubit steering inequality given in Eq. (\ref{new_steering1}) is violated, the other inequality given in Eq. (\ref{new_steering2}) cannot be violated and vice-versa. 

Although, here we have shown only one steering complementarity relation as in Eq. (\ref{comp2}), one can in principle derive a set of similar three qubit steering complementarity inequalities. One can also generalize the idea in the multi-qubit scenarios following the similar method.

\section{Conclusion}
In summary, we generalize the idea proposed in \cite{deba1} to establish a connection between two important resources of quantum information theory, namely quantum coherence and steering and provide a set of new coherence steering criteria. We classify these criteria in two classes for two-qubit states based on the number of measurements performed by Bob for each projective measurements performed by Alice. At last, we derive a set of complementarity relations between various coherence steering criteria. 

We also extended the work for three-qubit states and based on the two-qubit steering complementarity relations, we derived a set of new three-qubit steering inequalities, which can be generalized further for higher dimensions. Following the two-qubit derivations, we established the three-qubit steering complementarity relations.  These complementarity relations imply that if a state achieves NAQC due to one particular steering setting, it cannot achieve the same by its complementary setting.

It will be interesting to observe whether one can use such an advantage in coherence  in cryptographic protocols and other information theoretic tasks. One advantage of our relations is that quantum coherence is easily measurable quantity in the experiments. Studying its other advantages over the existing protocols will be off utmost importance in quantum cryptography, quantum communications and quantum information theory in general.

Moreover, we know that quantum coherence plays an important role in quantum information theory and quantum thermodynamics in setting the speed of quantum engines \cite{deba4}. It will interesting to study and understand the role of quantum steering as a resource in quantum thermodynamics in controlling thermodynamic engines nonlocally. We believe that this work will help to address such issues in quantum thermodynamics as well as quantum technologies in general in the future.

\section{Acknowledgements}
DM would like to acknowledge the support from
the National Research Foundation. DK is supported by the National Research Foundation and the Ministry of Education in Singapore through the Tier 3 MOE2012-T3-1-009 Grant ``Random numbers from quantum processes".

\end{document}